\documentclass[journal=ancac3,manuscript=article]{achemso}

\usepackage[version=3]{mhchem} 
\usepackage{geometry}
\usepackage{graphicx} 
\usepackage{amsmath,amssymb,xcolor}
\usepackage{braket,mathpazo,hyperref}
\usepackage{amsfonts}
\usepackage{amsmath}
\usepackage{mathtools}
\usepackage{subfigure}
\usepackage{multirow}
\usepackage{xcolor}

\author{Jordan Alan Edwards}
\affiliation[Imperial]
{Department of Materials and Thomas Young Centre for Theory and Simulation of Materials, Imperial College London, SW7 2AZ, United Kingdom}
\author{Johannes Lischner}
\email{j.lischner@imperial.ac.uk}
\affiliation[Imperial]
{Department of Materials and Thomas Young Centre for Theory and Simulation of Materials, Imperial College London, SW7 2AZ, United Kingdom}

\title{Material dependence of direct light-induced charge injection from metallic nanoparticles into adsorbates}


\begin{document}


\begin{tocentry}
  \includegraphics[width=9cm, height=3.5cm, keepaspectratio]{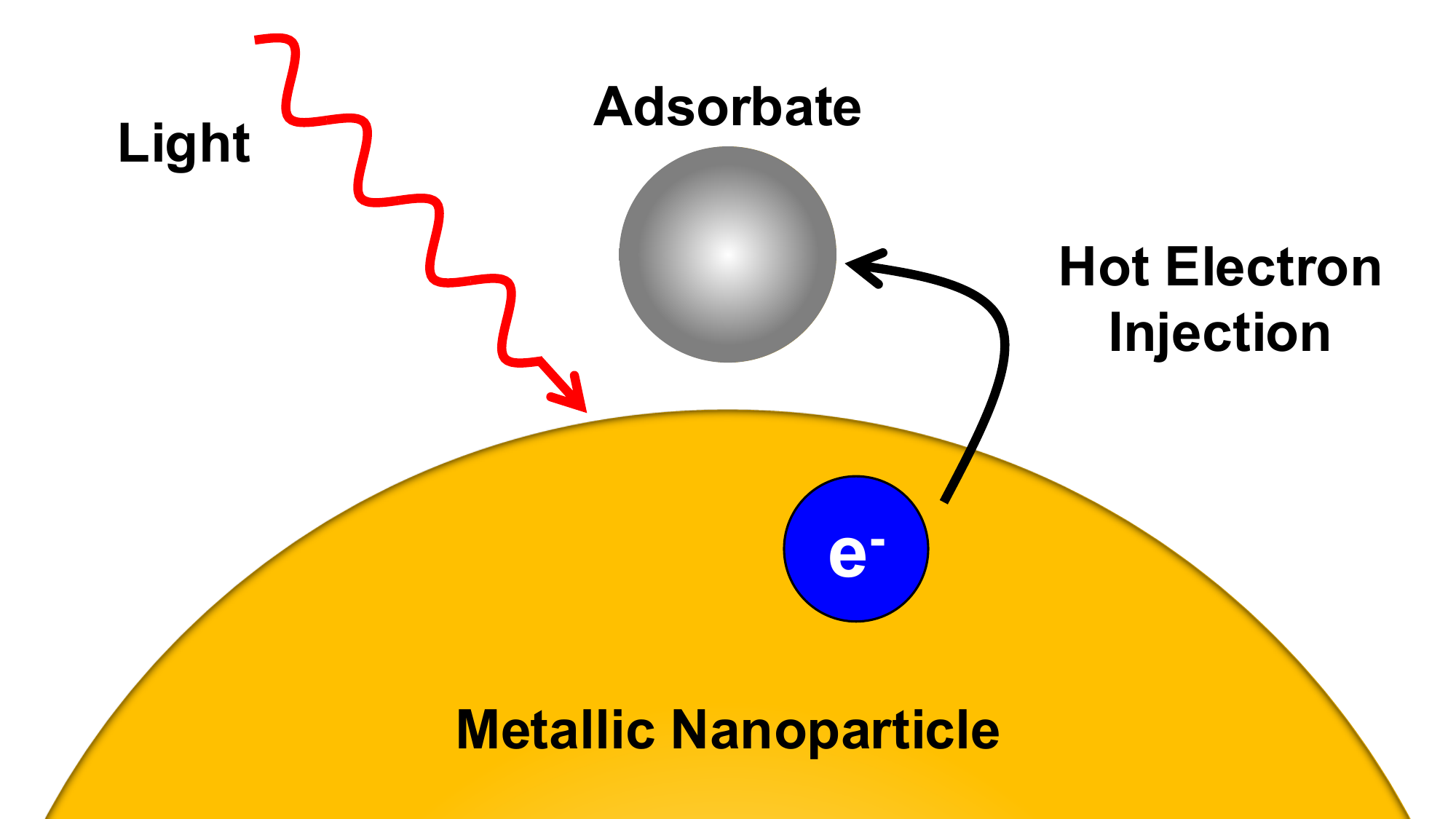}
\end{tocentry}

\begin{abstract}
Direct injection of charge from metallic nanoparticles into adsorbed molecules via the chemical interface damping of localized surface plasmons has emerged as a promising avenue to avoid the rapid thermalization of hot carriers. In this paper, we develop an approach for predicting hot-electron and hot-hole injection rates from large spherical nanoparticles into adsorbed carbon monoxide (CO) molecules. The approach combines a detailed description of the adsorbate-metal interactions via ab initio density functional theory with a macroscopic electrodynamics treatment of the localized surface plasmon. We apply this approach to nine different metals including both traditional (Ag, Au, Cu) and emerging (Al) plasmonic metals as well as non-plasmonic metals (Ni, Rh, Pd, Ir, Pt). For all metals, hot-electron injection is larger than hot-hole injection. This is a consequence of the alignment of the CO frontier orbitals, with the CO lowest unoccupied molecular orbital (LUMO) being closer to the metal Fermi energy than the highest occupied molecular orbital (HOMO). The largest electron injection rates in the optical photon ranges are obtained for Al, Ag, Cu and Au, with Cu and Au showing significant plasmonic enhancement. The non-plasmonic metals show weak electron injection as a consequence of their high LUMO energies relative to the metal Fermi level. The results of our study pave the way towards a rational design of metallic nanoscale photocatalysts for specific chemical processes.
\end{abstract}

\section{Introduction}
\label{sec:introduction}

In the last decade, there has been a growing interest in using metallic nanoparticles as photocatalysts~\cite{plasmonic-transformation, plasmonic-transformation-1, plasmonic-transformation-2, plasmonic-transformation-3, plasmonic-transformation-4, plasmonic-transformation-5, plasmonic-transformation-6, plasmonic-transformation-7}. Motivated by the large light-absorption cross-sections caused by localized surface plasmons (LSPs) in nanoparticles of Au and Ag, these efforts attempted to harness energetic or "hot" electrons and holes generated by the Landau decay of LSPs for inducing chemical transformations of adsorbed molecules. However, the energy conversion efficiencies of such photocatalytic devices are limited by the rapid thermalization of hot carriers which often lose a significant fraction of their energy before reaching the nanoparticle surface~\cite{thermalization-1, thermalization-2, thermalization-3, thermalization-4, thermalization-5,thermalization-6}.

This limitation can be overcome by exploiting the phenomenon of chemical interface damping (CID) of plasmons~\cite{CID1,CID2,CID3,CID4,CID5}. Here, the LSP decay involves adsorbate orbitals and therefore generates hot electrons or holes directly on the adsorbate. For example, Kale and coworkers demonstrated that CO oxidation on small Pt nanoparticles reaches a maximum at wavelengths at which a direct activation of the metal-adsorbate bond occurs via CID~\cite{kale}. Recently, Stefancu et al. coated Au waveguides with different molecules and demonstrated that the strength of CID is closely related to the alignment of the adsorbate frontier orbitals relative to the metal Fermi level~\cite{our-paper}. 

Important insights into CID have been provided by theoretical modelling techniques. A widely used model was developed by Persson~\cite{persson-polarisability} who evaluated Fermi's golden rule for the LSP decay and predicted an increase in the LSP linewidth of spherical metal nanoparticles due to the presence of adsorbates. Recently, this model was combined with ab initio density-functional theory calculations by Stefancu et al.~\cite{our-paper} following similar steps as Yuan and coworkers to model CID in TiO$_2$ coated Au nanorods~\cite{gold-tio2-persson-2-link}. For small nanoparticles, it is also possible to perform ab initio time-dependent density-functional theory calculations. For example, Kumar and coworkers studied direct light-induced charge injection into CO adsorbates on a 147-atom Ag nanoparticle~\cite{kumar}. Similar work was performed by Fojt and coworkers~\cite{fojt}. 

While much work has been focused of nanoparticles of Au and Ag, there is also an increasing effort to identify cheaper earth-abundant metals that can be used in nanoparticle photocatalysts~\cite{non-noble-1, non-noble-2, non-noble-3}. For example, Al, Cu and Mg nanoparticles exhibit LSPs and can exhibit photocatalytic activity~\cite{al-plasmonic, cu-plasmonic,mg-plasmonic}. Besides plasmonic metals, it has also become clear that nanoparticles of so-called "catalytic" metals, such as Pt or Pd, have interesting optical properties and can be used as photocatalysts. For example, a recent study by Henrotte et al. demonstrated that Pt nanoparticles produce more hot carriers than Au nanoparticles in certain wavelength regimes~\cite{naldoni-paper}. However, it is currently not clear how efficiently these materials facilitate direct hot-carrier injection into adsorbates via CID. 

To address this question, we present a theoretical study of direct hot-carrier injection from spherical nanoparticles into adsorbed carbon monoxide (CO). In particular, we extend the approach of Persson for calculating LSP linewidths~\cite{persson-polarisability} to yield hot-electron and hot-hole injection rates and perform calculations for the following metals: Al, Ag, Cu, Ni, Au, Rh, Pd, Ir and Pt, including both classical and emerging plasmonic metals as well as non-plasmonic "catalytic" metals. We find that all metals exhibit larger electron injection rates than hole injection rates. This is a consequence of the electronic structure of the CO-metal systems with the CO LUMO being much closer to the Fermi level than the CO HOMO level. The largest electron injection rates are found for the plasmonic metals (Al, Ag, Au and Cu) with Au and Cu demonstrating significant plasmonic enhancements in the optical regime.

\section{Results and discussion}
\label{sec:Results}

\subsection*{Electronic structure of CO on metal surfaces}

\label{subsec:pdos}
\begin{figure}[h!]
    \centering
    \includegraphics[width=\textwidth]{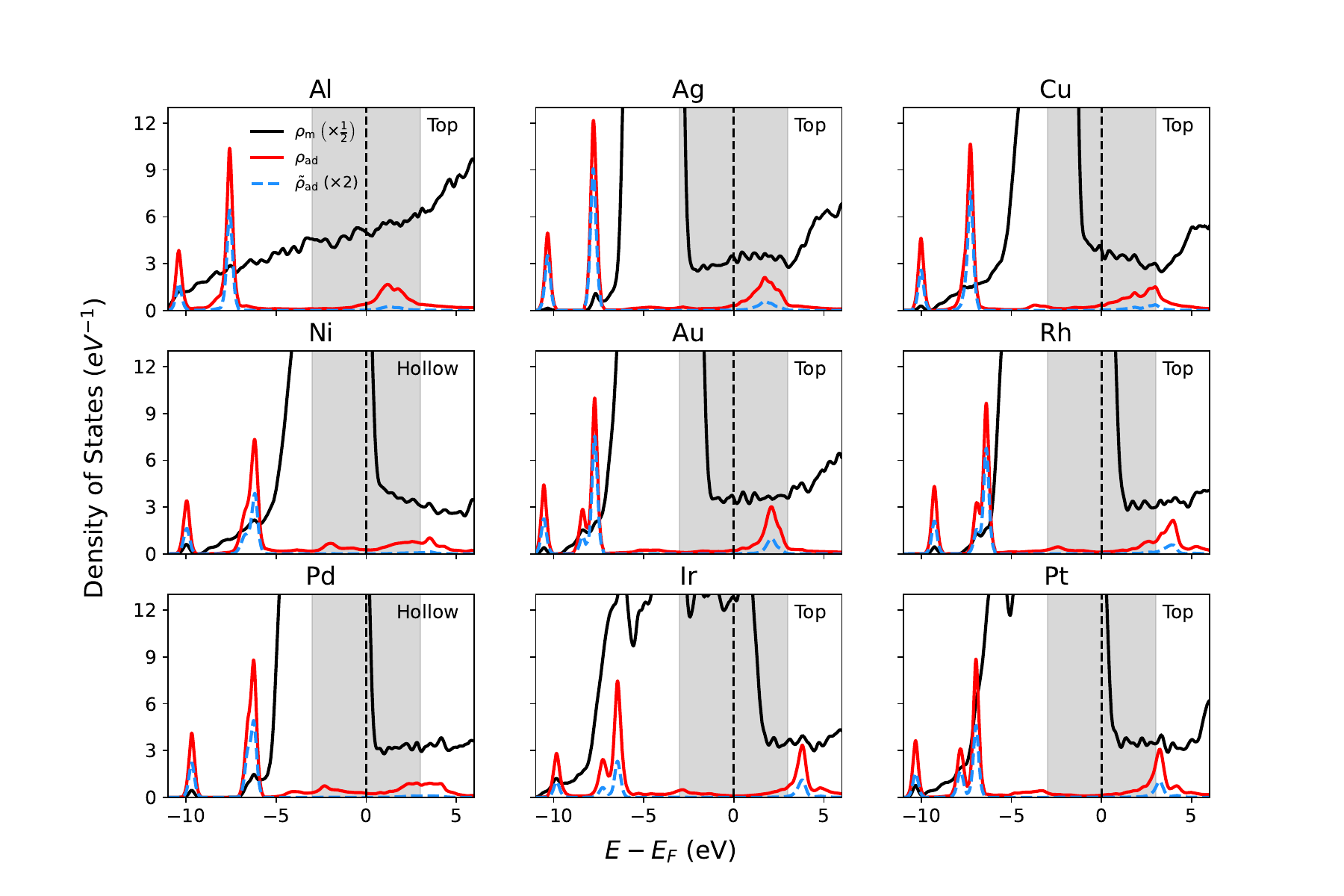}
    \caption{Density of states projected onto the CO orbitals $\rho_{\text{ad}}(\varepsilon)$ (solid red lines) and $\tilde{\rho}_{\text{ad}}(\varepsilon)$ (dashed blue lines) for (111) surfaces of different metals. The bulk metal density of states, $\rho_{\text{m}}(\varepsilon)$ is overlaid (solid black line). The dashed black line represents the Fermi level which has been set to $0\,\text{eV}$. Note that $\tilde{\rho}_{\text{ad}}$ has been multiplied by a factor of $5$, and $\rho_{\text{m}}(\varepsilon)$ by a factor of $0.5$. The gray area between $-3$ and $3\,\text{eV}$ highlights the range of photon energies for which charge injection is studied in Fig.~\ref{fig:ci-rate}. The experimental adsorption sites are stated in the top right corners.}
    \label{fig:co-pdos}
\end{figure}

Figure~\ref{fig:co-pdos} shows the adsorbate-projected density of states for CO on different (111) metal surfaces: Al, Ag, Cu, Ni, Au, Rh, Pd, Ir, and Pt. The panels are arranged according to the metal work function with lowest work function material (Al) on the top left and the highest (Pt) at the bottom right. 

For Al, Ag and Cu (top panels), CO adsorbs in the top position. For all systems, the adsorbate-projected DOS exhibits a peak above the Fermi level corresponding to the LUMO orbital of CO. Because of the hybridization with the metal states, the LUMO becomes a broadened resonance (centered at $1.2\,\text{eV}$ for Al, $1.7\,\text{eV}$ for Ag and approximately $2.2\,\text{eV}$ for Cu), with a full width at half maximum (FWHM) of $1.7\,\text{eV}$, $1.6\,\text{eV}$ and $2.2\,\text{eV}$ for Al, Ag and Cu respectively. The LUMO peaks are not fully symmetric: in particular for Ag and Cu these peaks exhibit a slow decay towards the Fermi level, but a much sharper rise at the high-energy side. In addition, the adsorbate DOS exhibits sharp peaks below the Fermi level corresponding to the HOMO resonances (centered at $-7.6\,\text{eV}$ for Al, $-7.8\,\text{eV}$ for Ag and $-7.3\,\text{eV}$ for Cu). The sharper HOMO features can be attributed to the lower density of metal states at the HOMO energy resulting in a weaker hybridization. Because the HOMO resonances are far from the Fermi level, we can neglect their contribution to charge injection induced by optical photons.

The adsorbate-projected DOS for Ni, Au and Rh are shown in the middle row of Fig.~\ref{fig:co-pdos}. For Ni, where CO adsorbs at the hollow site, the LUMO resonance, centered at approximately $2.8\,\text{eV}$, exhibits a broad plateau rather than a well-defined peak. Consequently, the FWHM of $2.7\,\text{eV}$ is significantly larger than those of Al, Ag and Cu. We find that for adsorption at hollow sites the adsorption distance, defined as the vertical separation between the topmost metal atom and the bottommost adsorbate atom, is significantly smaller than for top site adsorption (see Table 1 of the SI) allowing for stronger hybridization between adsorbate and metal states. Similar to Al, Ag and Au, Ni also shows a relatively sharp HOMO peak centered at $-6.2\,\text{eV}$. The adsorbate DOS for Au (top site adsorption) reveals a relatively symmetric LUMO feature centered at $2.1\,\text{eV}$ (FWHM of $0.9\,\text{eV}$), and a HOMO resonance centered at approximately $-7.7\,\text{eV}$. The adsorbate-projected DOS of Rh (top site adsorption) is qualitatively similar to that of Cu exhibiting a broad asymmetric LUMO resonance centered at $3.9\,\text{eV}$, with a FWHM of $1.1\,\text{eV}$.

Finally, the results for Pd, Ir and Pt are shown in the bottom row of Fig.~\ref{fig:co-pdos}. The adsorbate-projected DOS of Pd, on whose surface CO adsorbs at the hollow site, exhibits the same broad plateau-shaped LUMO resonance, centered at approximately $3\,\text{eV}$, as Ni. The adsorbate-projected DOS of Ir and Pt (both top site adsorption) exhibit similar shapes with relatively sharp and symmetric LUMO peak at $3.8\,\text{eV}$ (FWHM of $0.6\,\text{eV}$) for Ir and  $3.2\,\text{eV}$ (FWHM of $0.6\,\text{eV}$) for Pt.

\begin{figure}[t!]
    \centering
    \includegraphics[width=\textwidth]{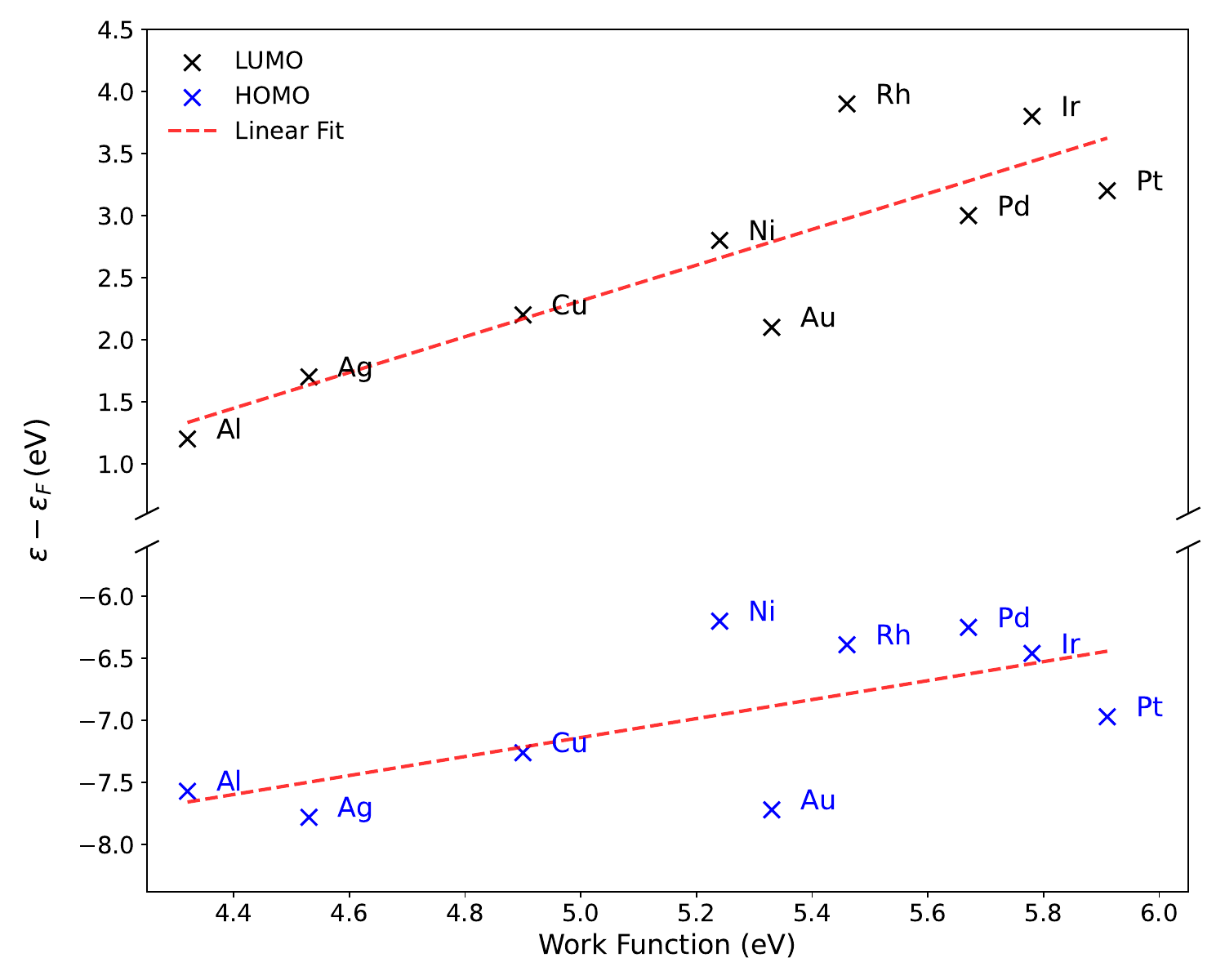}
    \caption{LUMO (black crosses) and HOMO (blue crosses) energies of CO on (111) surfaces of different metals obtained from DFT calculations versus the measured work functions. Dashed lines represent linear fits to the data.}
    \label{fig:workfunc-align}
\end{figure}

Figure~\ref{fig:workfunc-align} shows the energy of the HOMO and LUMO resonances (with respect to the Fermi level of the respective systems) from our calculations as function of the experimentally measured work functions of the different metals~\cite{workfunc-values}. Overall, we find a linear trend, i.e. high LUMO energies for metals with high work functions (and vice versa for the HOMO energies). Fitting both the HOMO and LUMO results to $y = mx + c$, we obtain $m = 1.44$ eV and $c = -4.89\,\text{eV}$ for the LUMO, and $m = 0.76$ eV and $c = -10.96\,\text{eV}$ for the HOMO. Several outliers are observed: in particular, the LUMO energy of Au is significantly lower than the linear fit, while that of Rh is significantly higher. Similarly, the HOMOs of Au and Ni show significant deviations from the linear behavior. This suggests that other quantities beyond the metal work function, such as d-band centers, also play a role in determining the HOMO and LUMO energy of CO.

Fig.~\ref{fig:co-pdos} also shows results for $\tilde{\rho}_{\text{ad}}(\varepsilon)$. We find that the shape of $\tilde{\rho}_{\text{ad}}(\varepsilon)$ is qualitatively similar to that of $\rho_{\text{ad}}(\varepsilon)$ for all metals. Quantitatively, we find that the higher power of the projection coefficient in the definition of $\tilde{\rho}_{\text{ad}}(\varepsilon)$ (see Eq.~\eqref{eq:rho-rho2-def}) suppresses broad peaks, see for example the CO LUMO resonance for Ni and Pd.

Previous studies of plasmon-induced charge injection\cite{lorentz-used-1, lorentz-used-2} often employed Lorentzian functions to describe the adsorbate-projected DOS. While such an ansatz can simplify subsequent calculations, our results reveal that the adsorbate-projected DOS often exhibit more complicated shapes with significant asymmetry. Moreover, for metals with hollow site adsorption the adsorbate-projected DOS does not even exhibit a clear LUMO peak. These results demonstrate that simplified analytical models of the adsorbate-projected DOS should be used with caution.


\subsection{Charge injection rates}
\label{subsec:charge-injection-rate}

\begin{figure}[h!]
    \centering
    \includegraphics[width=\textwidth]{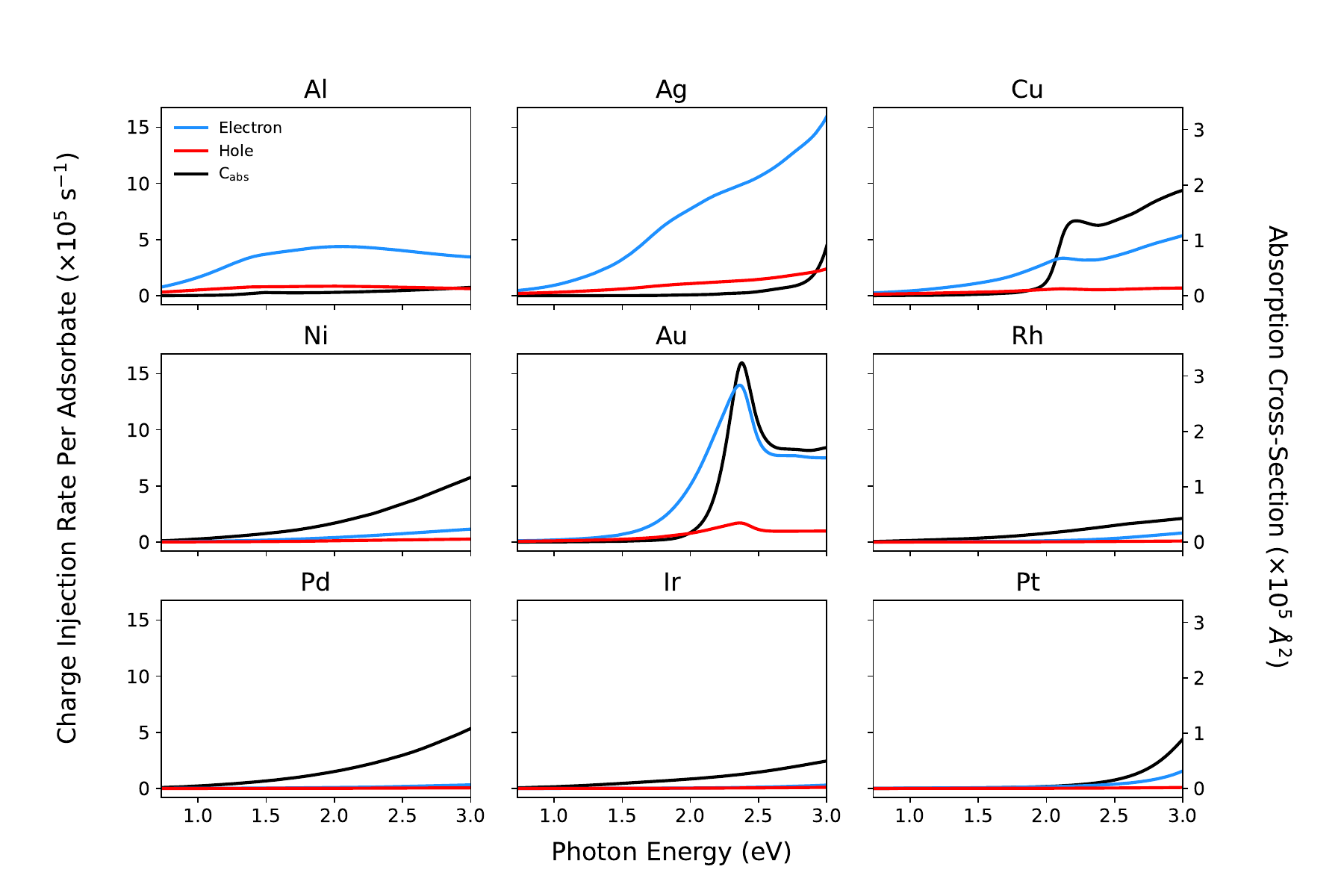}
    \caption{Electron (blue) and hole (red) injection rates per adsorbate for CO on spherical metal nanoparticles as a function of photon energy. Rates are evaluated for an electric field strength of $1000\,\text{V/m}$ (corresponding to approximately 1 solar intensity). The absorption cross section (black) for a nanoparticle of radius $R = 200\,\mathrm{\AA}$ is overlaid. }
    \label{fig:ci-rate}
\end{figure}

Figure~\ref{fig:ci-rate} shows the charge injection rates from metal nanoparticles (with a radius of $200\,\mathrm{\AA}$) into adsorbed CO molecules as a function of the photon energy. For all metals, we find that the electron injection rate is larger than hole injection rate. This is a consequence of the fact that the LUMO resonance of CO is much closer to the Fermi level than the HOMO resonance, see Fig.~\ref{fig:co-pdos}. Below, we focus our discussion therefore on the electron injection rates.

The top row of Fig.~\ref{fig:ci-rate} shows the charge injection rates for Al, Ag and Cu. For Al, the electron injection rate exhibits a broad maximum around $2\,\text{eV}$, corresponding to approximately the high-energy side of the LUMO resonance. The absorption cross-section of the Al nanoparticle exhibits a peak associated with the localized surface plasmon in the deep ultraviolet regime~\cite{al-plasmon-1, al-plasmon-2, al-plasmon-3}, but is quite small over the photon energy range considered in this work. As a consequence, charge injection in Al nanoparticles does not benefit from a plasmonic enhancement for optical photon energies. Despite this, Al exhibits a large electron injection rate compared to other metals.

For Ag, we observe a significant rise in the electron injection rate at low photon energies due to the onset of the LUMO resonance. The localized surface plasmon energy in this system is 3.5 eV corresponding to the onset in the absorption cross-section at 2.9 eV. As a consequence, charge injection only benefits from plasmonic enhancement at photon energies larger than 2.5 eV explaining the increase in the Ag's electron injection rate at high photon energies. 

For Cu, the electron injection rate exhibits a maximum near $2.2\,\text{eV}$. This peak coincides with the peak in the absorption cross-section $C_{\text{abs}}(\omega)$ which corresponds to the localized surface plasmon of the spherical Cu nanoparticle. When the plasmon condition $\epsilon(\omega_{pl})+2\epsilon_0=0$ is fulfilled, the electric field which drives charge injection into the adsorbate is dramatically enhanced, see Eq.~\eqref{eq:elec-inside}. At photon energies below the plasmon peak, electron injection rate rises more rapidly than $C_{abs}(\omega)$. This is caused by electron injection into the LUMO resonance which is not plasmonically enhanced.

Considering the middle row of panels, we find that the charge injection rates for Au nanoparticles exhibit a clear peak at $2.4\,\text{eV}$ which coincides with the localized surface plasmon peak of the absorption cross-section. Compared with Cu, the plasmon peak in $C_{\text{abs}}(\omega)$ is much more pronounced, and the stronger plasmonic field enhancement gives rise to a large peak in the electron injection rate. At higher photon energies, both the absorption cross-section and the charge injection rates exhibit a plateau. 

All other metals (Pt, Pd, Ir, Rh and Ni) exhibit comparatively small electron injection rates. For these metals, the energy of the LUMO resonance is too high to allow significant electron injection within the considered photon energy range and they also do not exhibit strong plasmonic field enhancement. Among these metals, the largest electron injection rate is obtained for Ni which exhibits a very broad LUMO resonance in addition to an increasing absorption cross-section.


\section{Conclusion}
We have studied the direct light-induced injection of charge from spherical metallic nanoparticles into adsorbed carbon monoxide using an approach that combines an ab initio description of the metal-adsorbate interaction with a macroscopic treatment of the localized surface plasmon. We considered nine different metals including both traditional (Ag, Au, Cu) and emerging (Al) plasmonic metals as well as non-plasmonic metals (Ni, Rh, Pd, Ir, Pt). For all metals, electron injection is larger than hole injection because the CO LUMO level always lies closer to the metal Fermi level than the CO HOMO level. The largest electron injection rates are obtained for Al, Ag, Cu and Au with Cu and Au showing a clear plasmonic enhancement in the optical regime. In contrast, the non-plasmonic metals Ni, Rh, Pd, Ir and Pt exhibit low electron injection rates as a consequence of the high CO LUMO energy (relative to the metal Fermi level). Our approach paves the way towards the computational design of metallic nanoparticle photocatalysts for the activation of specific metal-adsorbates bonds.

\section*{Methods}
\label{sec:methods}

\subsection{Rate of Charge Injection Into Adsorbates on Metal Nanoparticles}
\label{subsec:theory-charge-injection}

Following Persson~\cite{persson-polarisability,persson-work-func}, we consider a spherical metallic nanoparticle with a single molecule adsorbed to its surface. For simplicity, we assume that the adsorbate has one orbital $| n \rangle$. If this system is illuminated by light of frequency $\omega$, the rate of direct electron injection into the adsorbate is given by

\begin{equation}
    C_e(\omega) = \sum_{\alpha\beta}N^{\text{ad}}_{\beta}\Gamma_{\alpha\beta}(\omega),
    \label{eq:elec-injec-rate}
\end{equation}
where $N^{\text{ad}}_{\beta}= |\langle n | \beta \rangle|^2$ denotes the fraction of the final state that resides on the adsorbate and $\Gamma_{\alpha\beta}(\omega)$ is the transition rate from initial state $| \alpha \rangle$ to final state $|\beta \rangle $ given by Fermi's golden rule as  

\begin{equation}
    \Gamma_{\alpha\beta}(\omega) = \frac{2\pi}{\hbar}f_\alpha(1-f_\beta)|\braket{\beta|\hat H_{int}|\alpha}|^2\delta(\varepsilon_\beta - \varepsilon_\alpha - \hbar\omega),
    \label{eq:fermi-original}
\end{equation}
where $\hbar$ is the reduced Planck constant, $\varepsilon_\alpha$ and $f_\alpha$ denote the energy and ground-state occupancy of state $\ket{\alpha}$. Also, $\hat H_{int}$ is the interaction Hamiltonian which is given by~\cite{persson-work-func}
\begin{equation}
    \hat H_{int}(t) = -\hat \mu \cdot \vec{E}(t),
    \label{eq:interaction-hamiltonian}
\end{equation}
where $\hat \mu$ is the dipole moment operator and $\vec{E}(t)$ is the time-dependent electric field. For charge-transfer excitations between the adsorbate and the nanoparticle, we assume that only the radial component of the transition dipole is non-zero and the corresponding matrix element is given by~\cite{persson-work-func}

\begin{equation}
    \langle \alpha |\hat \mu_r | \beta \rangle = -e d \langle \alpha | n \rangle \langle n | \beta \rangle,
    \label{eq:dipmom-operator}
\end{equation}
where $e$ is the proton charge, $d$ is the distance of the adsorbate to the nanoparticle surface and $\hat r$ is the radial unit vector on the surface of the spherical nanoparticle. Note that we neglected the contribution to the dipole moment operator from the electronic polarizability.

Within the quasi-static approximation, the electric field of a spherical metallic nanoparticle with radius $R$ illuminated by light of frequency $\omega$ is given by $\vec{E}(t) = \vec{E}(|\vec{r}|, \theta, \omega) e^{-i\omega t}$ with

\begin{align}
\vec{E}(|\vec{r}|, \theta, \omega) = &E_0[\cos(\theta)\hat{r} - \sin(\theta)\hat{\theta}] \notag\\ 
+ &E_0 \frac{\epsilon(\omega) - \epsilon_0}{\epsilon(\omega) + 2\epsilon_0} \left(\frac{R}{|\vec{r}|}\right)^3\left[2\cos(\theta)\hat{r} - \sin(\theta)\hat{\theta}\right], \quad |\vec{r}| > R,
\label{eq:elec-outside}
\end{align}

\begin{align}
\vec{E}(|\vec{r}|, \theta, \omega) =\frac{3\epsilon_0}{\epsilon(\omega) + 2\epsilon_0}E_0\left[\cos(\theta)\hat{r} - \sin(\theta)\hat{\theta}\right], \quad |\vec{r}| < R,
\label{eq:elec-inside}
\end{align}
where $\epsilon_0$ is the dielectric constant of free space, $\epsilon(\omega)$ is the frequency-dependent dielectric function of the metal, $E_0$ is the electric field strength, $|\vec{r}|$ is the distance from the center of the nanoparticle, and $\theta$ is the angle between the incident electric field and the surface normal $\hat{r}$ (with associated basis vector $\hat \theta$). Note that for a spherical nanoparticle $\hat{\mu}$ is parallel to $\hat{r}$ and therefore only couples to the radial component of the electric field. In our calculations, we approximate the electric field at the position of the adsorbate by $\vec{E}(R, \theta, \omega)$.

Inserting the expressions for the electric field and the dipole moment operator into Eq.~\eqref{eq:elec-injec-rate} yields
\begin{equation}
    C_e(\omega) = A(\omega)\sum_{\alpha\beta}f_\alpha(1-f_\beta) \langle \beta| n \rangle |^4 |\langle \alpha|n \rangle|^2  \delta(\varepsilon_\beta - \varepsilon_\alpha - \hbar\omega),
\end{equation}
where $A(\omega)$ is given by

\begin{equation*}
    A(\omega) = \frac{2\pi}{\hbar}\left|\frac{3\epsilon(\omega)}{\epsilon(\omega) + 2\epsilon_0}\right|^2E_0^2(ed)^2 \cos^2(\theta).
\end{equation*}
Following Dubois and Persson~\cite{persson-work-func}, this expression can be re-written as

\begin{equation}
    C_e(\omega) = A(\omega)\int_{\varepsilon_F - \hbar\omega}^{\varepsilon_F}d\varepsilon \rho_{\text{ad}}(\varepsilon)\tilde{\rho}_{\text{ad}}(\varepsilon + \hbar\omega),
    \label{eq:c-e}
\end{equation}
where we introduced the density of states (DOS) projected onto adsorbate orbitals $\rho_{\text{ad}}(\varepsilon)$ and the adsorbate-charge weighted DOS projected onto the adsorbate $\tilde{\rho}_{\text{ad}}(\varepsilon)$ given by
\begin{align}
    \rho_{\text{ad}}(\varepsilon) &= \sum_{\alpha}|\braket{\alpha|n}|^2\delta(\varepsilon - \varepsilon_\alpha),\\
    \tilde{\rho}_{\text{ad}}(\varepsilon) &= \sum_{\alpha}|\braket{\alpha|n}|^4\delta(\varepsilon - \varepsilon_\alpha).
    \label{eq:rho-rho2-def}
\end{align}

Following similar steps, the rate of hole injection into the adsorbate can be obtained as
\begin{equation}
    C_h(\omega) = A(\omega)\int_{\varepsilon_F - \hbar\omega}^{\varepsilon_F}d\varepsilon \tilde{\rho}_{\text{ad}}(\varepsilon)\rho_{\text{ad}}(\varepsilon+\hbar\omega) .
    \label{eq:c-h}
\end{equation}

To describe a nanoparticle with a uniform coverage of adsorbates, we average the charge injection rate over the adsorbate position resulting in the replacement of $\braket{\cos^2\theta} \rightarrow \frac{1}{3}$.

\subsection{Computational Details}
\label{subsec:computational Details}

To evaluate the electron and hole injection rates, various quantities, such as the adsorbate-projected density of states, are required. To obtain these, we perform ab initio DFT calculations of metal surfaces with adsorbed CO molecules. Specifically, we study the (111) surfaces of Al, Ag, Cu, Ni, Au, Rh, Pd, Ir, and Pt using a periodic slab model built from the DFT-optimized bulk lattice parameters. We create a $2\times2$ supercell and add the CO molecule at the experimental adsorption position reported in Refs.~\cite{co-adsorption-1, co-adsorption-2, co-adsorption-3} (see Table 1 of the Supporting Information). This corresponds to the low-coverage limit ($\Theta = 1/4$ monolayer). The slabs have six atomic layers, a thickness at which the surface energy was converged, and are separated by $20\,\text{\AA}$ of vacuum to prevent interaction between periodically repeated images.

All calculations were performed using the Quantum Espresso software package with a GGA-PBE exchange-correlation functional and norm-conserving pseudopotentials~\cite{qe1,qe2,qe3,pbe-func,pseudodojo1, pseudodojo2, pseudodojo3}. Van der Waals interactions were accounted for using the Grimme-D3 correction and a dipole correction was applied ~\cite{grimme-d3,dipole-correction}. We used a plane-wave cutoff of $1292\, \text{eV}$ and a $6$x$6$x$1$ Monkhorst-Pack $k$-point grid for sampling the first Brillouin zone. An energy convergence criterion of $10^{-8}\, \text{eV}$ was used in the self-consistent cycles with a maximum force tolerance of $0.003\, \text{eV/\AA}$. Spin polarization was employed for Ni and Pd (which has been found to be a ferromagnet within DFT-PBE~\cite{pd-ferro-1, pd-ferro-2}). 
For the calculation of the adsorbate-projected DOS, the Kohn-Sham eigenstates are projected onto localized atomic-like pseudo-orbitals associated with the relevant pseudopotentials. Finally, the dielectric functions of the metals are taken from experimental measurements~\cite{dielec-1, dielec-2, dielec-3, dielec-4, dielec-5, dielec-6, dielec-7, dielec-8} and $d$ was set to the distance between CO and the metal surface obtained from the DFT relaxations, see Table 1 of the Supplementary Information.

\begin{acknowledgement}
The authors thank Emiliano Cortes and Rupert Oulton for useful discussions.
JAE acknowledges support from the Engineering and Physical Sciences Research Council (grant number grant number EP/W524323/1), and also from the EPSRC program grant EP/W017075/1. JL acknowledges support from a Miller Visiting Professorship at UC Berkeley. Via our membership of the UK's HEC Materials Chemistry Consortium (MCC), which is funded by EPSRC (EP/X035859), this work used the ARCHER2 UK National Supercomputing Service (\href{https://www.archer2.ac.uk}{https://www.archer2.ac.uk}) and the Imperial College Research Computing Service facility CX1 (DOI: 10.14469/hpc/2232).

\end{acknowledgement}

\begin{suppinfo}
The Supporting Information is available free of charge.

Convergence parameters for the bulk metals and slab systems and
parameters used to evaluate the
charge injection rate
\end{suppinfo}


\clearpage

\begin{center}
    {\LARGE\sffamily\bfseries Supplementary information for material dependence of direct light-induced charge injection from metallic nanoparticles into adsorbates \par}
    \vspace{1.5em}
 
    {\large\sffamily Jordan Alan Edwards, Johannes Lischner$^{*}$\par}
    \vspace{1em}
    
    {\small\itshape 
    Department of Materials and Thomas Young Centre for Theory and Simulation of Materials,\\ Imperial College London, SW7 2AZ, United Kingdom \par}
    \vspace{0.8em}
    
    {\small\sffamily E-mail: j.lischner@imperial.ac.uk \par}
    \vspace{2em}
\end{center}

\section{Computational parameters for bulk metals}
\label{sec:bulk}

To determine convergence parameters for metal slabs, we first studied bulk metals. For these, we first determine the required plane-wave cutoff energy, then the $k$-point grid, and finally we obtain the lattice parameter. The plane-wave cutoff energy for the wavefunction expansion was first converged. The cutoff was increased until the total energy per electron changed by less than $10 \,\mathrm{meV}$ relative to the most accurate calculation, which used a cutoff of approximately $2040 \,\mathrm{eV}$. A Fermi-Dirac smearing with a broadening of $0.07\,\mathrm{eV}$ was used in all calculations. The largest plane-wave cutoff was used in all subsequent calculations, which was found to be $1292\,\mathrm{eV}$ for both Cu and Ni.
Using the converged plane-wave cutoff, the $k$-point grid was varied until the until the total energy per electron changed by less than $10 \,\mathrm{meV}$ relative to the densest grid ($30\times30\times30$). The finest $k$-point grids ($12\times12\times12$) were required for Au, Ag, Pt and Rh, which again was used in all subsequent bulk calculations.
Finally, the lattice parameter was optimized by scanning $\pm 3\%$ around the experimental value and computing the total energy as a function of lattice parameter. A parabola was fit to the resulting total energy curve, and its minimum was taken as the equilibrium lattice parameter. This procedure was repeated, with each new scan centered on the previously obtained minimum, until successive lattice parameters differed by less than $0.01\,\mathrm{\AA}$. The converged lattice parameters were then used to construct the slabs.


\section{Slab Convergence}
\label{sec:slab-conv}

All of the metals studied are FCC, for which the (111) facet typically has the lowest surface energy~\cite{all-bulk-surface}. This surface was therefore used throughout.
Slabs were constructed using the DFT-derived lattice parameters as periodic $1 \times 1 \times N$ supercells, where $N$ is the number of atomic layers along the surface normal ($z$). A vacuum region of $20\,\mathrm{\AA}$ between periodic images (i.e. $10\,\mathrm{\AA}$ either side of the slab) was used.

The number of layers $N$ was converged by calculating the surface energy $\sigma$ of $1 \times 1 \times N$ slabs using the method of Fiorentini and Methfessel~\cite{fiorentini}, with the plane-wave cutoff and $k$-point sampling determined from the bulk calculations. These calculations were static total-energy calculations on unrelaxed slabs. The surface energies were found to converge at around $6$ layers.


\section{CO Adsorption}

\begin{figure}[h!]
    \centering
    \includegraphics[width=\textwidth]{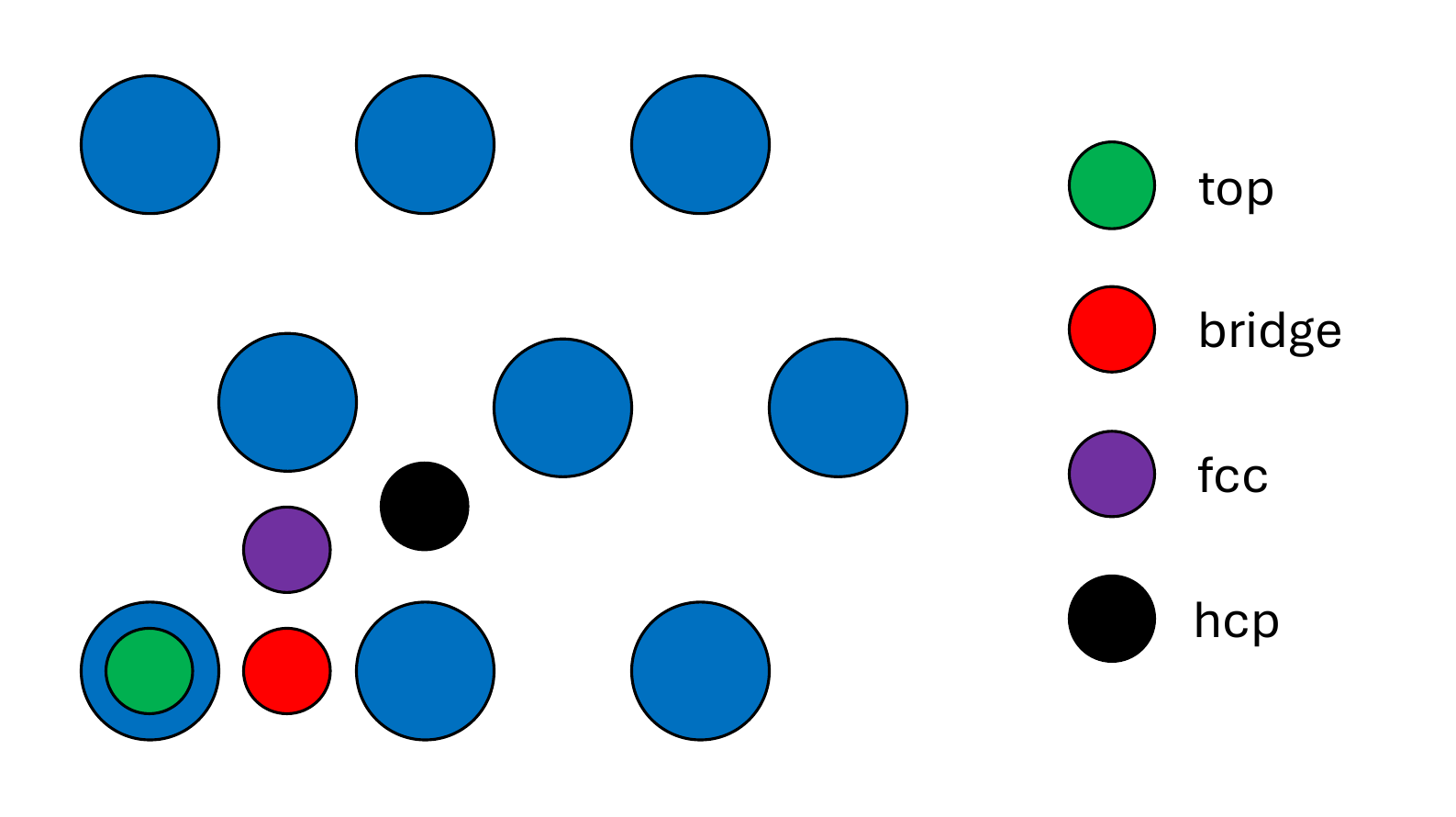}
    \caption{The FCC (111) surface (blue) has four non-equivalent adsorption sites. These relate to the coordination of each site and are denoted as top (green), bridge (red), fcc (purple) and hcp (black).}
    \label{fig:111-ads-site}
\end{figure}

The preferred adsorption site and distance of CO on each ($111$) surface was investigated through a series of geometry optimizations. A $2\times2$ supercell was constructed with a single CO molecule, corresponding to a coverage of $\Theta = 1/4$ monolayer (one adsorbate per 4 surface metal atoms). The CO molecule was placed at each of the four high-symmetry sites shown in Fig.~\ref{fig:111-ads-site}: bridge (bri), face centered cubic hollow (fcc), hexagonal close-packed hollow (hcp) and atop (top). CO was initially oriented with the C atom closest to the metal surface at an adsorption height of $2.5\,\mathrm{\AA}$. To break the symmetry of the initial forces, the molecule was tilted by $35^\circ$ relative to the surface normal ($z$). All calculations used the plane-wave cutoff, smearing and spin-polarization settings from the bulk and adsorbate-free slab convergence tests. A dipole correction was applied along the $z$-direction and the Grimme-D3 dispersion correction was used \cite{dipole-correction,grimme-d3}. All atoms of the slab and the adsorbate were allowed to relax during the optimization until a maximum force tolerance of $0.003\, \text{eV/\AA}$ was achieved. The adsorption distance is defined as the vertical distance (along $z$) between the topmost metal atom and the C atom of the adsorbed CO. The final relaxed geometry of each starting configuration was inspected, and Table~\ref{tab:co-ads-dist} reports the adsorption distances only for the experimentally observed adsorption site.

\begin{table}
    \centering
    \begin{tabular}{c c c}
    \hline
    \hline
    Metal & Adsorption Distance (\AA) & Experimental Adsorption Site~\cite{co-adsorption-ag, co-adsorption-pt, co-adsorption-pd, co-adsorption-ir, co-adsorption-rh, co-adsorption-cu, co-adsorption-dft1, co-adsorption-dft2, co-adsorption-dft3}\\
    \hline \vspace{-5mm} \\
    Au & 2.00 & Top\\
    
    \hline
    Ag & 2.12 & Top\\

    \hline
    Al & 2.02 & Top\\

    \hline
    Cu & 1.84 & Top\\

    \hline
    Pt & 1.84 & Top\\

    \hline
    Pd & 1.31 & FCC/HCP\\

    \hline
    Ir & 1.84 & Top\\

    \hline
    Rh & 1.84 & Top\\

    \hline
    Ni & 1.32 & FCC/HCP\\
    
    \hline
    \hline
    \end{tabular}
    \caption{Adsorption distance of CO at low coverages ($\Theta = 1/4$) on (111) metal surfaces in the relaxed geometry.}
    \label{tab:co-ads-dist}
\end{table}

\bibliography{acs-achemso.bib}

@article{persson-polarisability,
  title={Polarizability of small spherical metal particles: influence of the matrix environment},
  author={Persson, BNJ},
  journal={Surface Science},
  volume={281},
  number={1-2},
  pages={153--162},
  year={1993},
  publisher={Elsevier}
}

@article{qe1,
  author={P Giannozzi and O Andreussi and T Brumme and O Bunau and M Buongiorno Nardelli
  and M Calandra and R Car and C Cavazzoni and D Ceresoli and M Cococcioni and N Colonna
  and I Carnimeo and A Dal Corso and S de Gironcoli and P Delugas and R A DiStasio Jr and A Ferretti
  and A Floris and G Fratesi and G Fugallo and R Gebauer and U Gerstmann and F Giustino and T Gorni
  and J Jia and M Kawamura and H-Y Ko and A Kokalj and E Küçükbenli and M Lazzeri and M Marsili
  and N Marzari and F Mauri and N L Nguyen and H-V Nguyen and A Otero-de-la-Roza and L Paulatto
  and S Poncé and D Rocca and R Sabatini and B Santra and M Schlipf and A P Seitsonen
  and A Smogunov and I Timrov and T Thonhauser and P Umari and N Vast and X Wu and S Baroni},
  title={Advanced capabilities for materials modelling with QUANTUM ESPRESSO},
  journal={Journal of Physics: Condensed Matter},
  volume={29},
  number={46},
  pages={465901},
  url={http://stacks.iop.org/0953-8984/29/i=46/a=465901},
  year={2017},
}

@article{qe2,
	Author = {Paolo Giannozzi and Stefano Baroni and Nicola Bonini and Matteo Calandra and Roberto Car
 and Carlo Cavazzoni and Davide Ceresoli and Guido L Chiarotti and Matteo Cococcioni and Ismaila Dabo
 and Andrea {Dal Corso} and Stefano de Gironcoli and Stefano Fabris and Guido Fratesi and Ralph Gebauer
 and Uwe Gerstmann and Christos Gougoussis and Anton Kokalj and Michele Lazzeri and Layla Martin-Samos
 and Nicola Marzari and Francesco Mauri and Riccardo Mazzarello and Stefano Paolini and Alfredo Pasquarello
 and Lorenzo Paulatto and Carlo Sbraccia and Sandro Scandolo and Gabriele Sclauzero and Ari P Seitsonen
 and Alexander Smogunov and Paolo Umari and Renata M Wentzcovitch},
	Journal = {Journal of Physics: Condensed Matter},
	Number = {39},
	Pages = {395502 (19pp)},
	Title = {QUANTUM ESPRESSO: a modular and open-source software project for quantum simulations of materials},
	Url = {http://www.quantum-espresso.org},
	Volume = {21},
	Year = {2009}}

@article{qe3,
 author = {Giannozzi,Paolo  and Baseggio,Oscar  and Bonfà,Pietro  and Brunato,Davide  and Car,Roberto  and Carnimeo,Ivan  
           and Cavazzoni,Carlo  and de Gironcoli,Stefano  and Delugas,Pietro  and Ferrari Ruffino,Fabrizio  and Ferretti,Andrea  
           and Marzari,Nicola  and Timrov,Iurii  and Urru,Andrea  and Baroni,Stefano },
 title = {Quantum ESPRESSO toward the exascale},
 journal = {The Journal of Chemical Physics},
 volume = {152},
 number = {15},
 pages = {154105},
 year = {2020},
 doi = {10.1063/5.0005082},
 
 URL = { 
        https://doi.org/10.1063/5.0005082
    },
eprint = { 
        https://doi.org/10.1063/5.0005082
         }
}

@article{pseudodojo1,
  title={The PseudoDojo: Training and grading a 85 element optimized norm-conserving pseudopotential table},
  author={Van Setten, Michiel J and Giantomassi, Matteo and Bousquet, Eric and Verstraete, Matthieu J and Hamann, Don R and Gonze, Xavier and Rignanese, G-M},
  journal={Computer Physics Communications},
  volume={226},
  pages={39--54},
  year={2018},
  publisher={Elsevier}
}

@article{pseudodojo2,
  title={Reproducibility in density functional theory calculations of solids},
  author={Lejaeghere, Kurt and Bihlmayer, Gustav and Bj{\"o}rkman, Torbj{\"o}rn and Blaha, Peter and Bl{\"u}gel, Stefan and Blum, Volker and Caliste, Damien and Castelli, Ivano E and Clark, Stewart J and Dal Corso, Andrea and others},
  journal={Science},
  volume={351},
  number={6280},
  pages={aad3000},
  year={2016},
  publisher={American association for the advancement of science}
}

@article{pseudodojo3,
  title={Optimized norm-conserving Vanderbilt pseudopotentials},
  author={Hamann, DR},
  journal={Physical Review B—Condensed Matter and Materials Physics},
  volume={88},
  number={8},
  pages={085117},
  year={2013},
  publisher={APS}
}

@article{grimme-d3,
  title={A consistent and accurate ab initio parametrization of density functional dispersion correction (DFT-D) for the 94 elements H-Pu},
  author={Grimme, Stefan and Antony, Jens and Ehrlich, Stephan and Krieg, Helge},
  journal={The Journal of chemical physics},
  volume={132},
  number={15},
  year={2010},
  publisher={AIP Publishing}
}

@article{dipole-correction,
  title={Dipole correction for surface supercell calculations},
  author={Lennart Bengtsson},
  journal={Physical review B},
  volume={59},
  number={12301},
  year={1999},
  publisher={AIP }
}

@article{pbe-func,
  title={Generalized gradient approximation made simple},
  author={Perdew, John P and Burke, Kieron and Ernzerhof, Matthias},
  journal={Physical review letters},
  volume={77},
  number={18},
  pages={3865},
  year={1996},
  publisher={APS}
}

@article{fiorentini,
  title={Extracting convergent surface energies from slab calculations},
  author={Fiorentini, Vincenzo and Methfessel, Michael},
  journal={Journal of Physics: Condensed Matter},
  volume={8},
  number={36},
  pages={6525},
  year={1996},
  publisher={IOP Publishing}
}

@article{co-adsorption-rh,
  title={Desorption and molecular interactions on surfaces: C) Rh (110), CORh (100) and CORh (111)},
  author={Wei, DH and Skelton, DC and Kevan, SD},
  journal={Surface science},
  volume={381},
  number={1},
  pages={49--64},
  year={1997},
  publisher={Elsevier}
}

@article{co-adsorption-ir,
  title={Interaction of CO and O2 with Ir (111) surfaces},
  author={K{\"u}ppers, J and Plagge, A},
  journal={Journal of Vacuum Science and Technology},
  volume={13},
  number={1},
  pages={259--263},
  year={1976},
  publisher={AIP Publishing}
}

@article{co-adsorption-pd,
  title={Adsorption and coadsorption of carbon monoxide and hydrogen on Pd (111)},
  author={Kiskinova, MP and Bliznakov, GM},
  journal={Surface Science},
  volume={123},
  number={1},
  pages={61--76},
  year={1982},
  publisher={Elsevier}
}

@article{co-adsorption-ag,
  title={Adsorption of carbon monoxide on silver/palladium alloys},
  author={Christmann, K and Ertl, G},
  journal={Surface Science},
  volume={33},
  number={2},
  pages={254--270},
  year={1972},
  publisher={Elsevier}
}

@article{co-adsorption-pt,
  title={Effects of K, O, and H adatoms on the adsorption kinetics of CO on Pt (111)},
  author={Jiang, LQ and Koel, BE and Falconer, JL},
  journal={Surface science},
  volume={273},
  number={3},
  pages={273--284},
  year={1992},
  publisher={Elsevier}
}

@article{co-adsorption-cu,
  title={Interactions of CO molecules adsorbed on Cu (111)},
  author={Hollins, P and Pritchard, J},
  journal={Surface Science},
  volume={89},
  number={1-3},
  pages={486--495},
  year={1979},
  publisher={Elsevier}
}

@article{co-adsorption-dft2,
  title={CO adsorption on close-packed transition and noble metal surfaces: trends from ab initiocalculations},
  author={Gajdo{\v{s}}, Marek and Eichler, Andreas and Hafner, J{\"u}rgen},
  journal={Journal of Physics: Condensed Matter},
  volume={16},
  number={8},
  pages={1141},
  year={2004},
  publisher={IOP Publishing}
}

@article{co-adsorption-dft1,
  title={CO adsorption energies on metals with correction for high coordination adsorption sites--A density functional study},
  author={Abild-Pedersen, Frank and Andersson, Martin P},
  journal={Surface Science},
  volume={601},
  number={7},
  pages={1747--1753},
  year={2007},
  publisher={Elsevier}
}

@article{co-adsorption-dft3,
  title={CO adsorption on metal surfaces: A hybrid functional study with plane-wave basis set},
  author={Stroppa, Alessandro and Termentzidis, Konstantinos and Paier, Joachim and Kresse, Georg and Hafner, J{\"u}rgen},
  journal={Physical Review B—Condensed Matter and Materials Physics},
  volume={76},
  number={19},
  pages={195440},
  year={2007},
  publisher={APS}
}

@article{our-paper,
  title={Optical and electrical probing of plasmonic metal-molecule interactions},
  author={Stefancu, Andrei and Tang, Wenxuan and Fu, Ming and Edwards, Jordan and Halas, Naomi J and Schofield, Ross C and Millard, Toby Severs and Nordlander, Peter and Lischner, Johannes and Carro, Pilar and others},
  journal={Science Advances},
  volume={11},
  number={50},
  pages={eaea0585},
  year={2025},
  publisher={American Association for the Advancement of Science}
}

@article{CID1,
  title={Plasmon damping depends on the chemical nature of the nanoparticle interface},
  author={Foerster, Benjamin and Spata, Vincent A and Carter, Emily A and S{\"o}nnichsen, Carsten and Link, Stephan},
  journal={Science advances},
  volume={5},
  number={3},
  pages={eaav0704},
  year={2019},
  publisher={American Association for the Advancement of Science}
}

@article{CID2,
  title={Chemical interface damping depends on electrons reaching the surface},
  author={Foerster, Benjamin and Joplin, Anneli and Kaefer, Katharina and Celiksoy, Sirin and Link, Stephan and Sönnichsen, Carsten},
  journal={ACS nano},
  volume={11},
  number={3},
  pages={2886--2893},
  year={2017},
  publisher={ACS Publications}
}

@article{CID3,
  title={Chemical interface damping of surface plasmon resonances},
  author={Lee, Stephen A and Link, Stephan},
  journal={Accounts of chemical research},
  volume={54},
  number={8},
  pages={1950--1960},
  year={2021},
  publisher={ACS Publications}
}

@article{CID4,
  author  = {Kim, Jae-Myoung and
             Chatterjee, Subhojyoti and
             Gomez, Eric and
             Mehmood, Arshad and
             Yuan, Tinglian and
             Levine, Benjamin G. and
             Link, Stephan and
             Landes, Christy F.},
  title   = {Electrochemically Tunable Chemical Interface Damping in
             Methyl Viologen-Modified Single Gold Nanorods},
  journal = {ACS Nano},
  year    = {2026},
  volume  = {20},
  number  = {35},
  pages   = {24580--24592},
  doi     = {10.1021/acsnano.6c11027},
  url     = {https://doi.org/10.1021/acsnano.6c11027}
}

@article{CID5,
  author  = {Pfeiffer, Maurice and
             Vonbun-Feldbauer, Gregor B. and
             Khurgin, Jacob B. and
             Matts, Olga and
             Shqer, Ahmed and
             Mameka, Nadiia and
             Eich, Manfred and
             Petrov, Alexander},
  title   = {Energy-Dependent Chemical Interface Damping Induced by
             1-Decanethiol Self-Assembled Monolayer on Au(111)},
  journal = {The Journal of Physical Chemistry C},
  year    = {2026},
  volume  = {130},
  number  = {8},
  pages   = {3096--3101},
  doi     = {10.1021/acs.jpcc.5c07634},
  url     = {https://doi.org/10.1021/acs.jpcc.5c07634}
}

@article{kumar,
  author  = {Kumar, P. V. and Rossi, T. P. and Kuisma, M. and
             Erhart, P. and Norris, D. J.},
  title   = {Direct hot-carrier transfer in plasmonic catalysis},
  journal = {Faraday Discussions},
  year    = {2019},
  volume  = {214},
  pages   = {189--197},
  doi     = {10.1039/C8FD00154E},
  url     = {https://doi.org/10.1039/C8FD00154E}
}

@article{fojt,
  author  = {Fojt, Jakub and Rossi, Tuomas P. and
             Kuisma, Mikael and Erhart, Paul},
  title   = {Hot-Carrier Transfer across a Nanoparticle--Molecule
             Junction: The Importance of Orbital Hybridization
             and Level Alignment},
  journal = {Nano Letters},
  year    = {2022},
  volume  = {22},
  number  = {21},
  pages   = {8786--8792},
  doi     = {10.1021/acs.nanolett.2c02327},
  url     = {https://doi.org/10.1021/acs.nanolett.2c02327}
}

@article{non-noble-1,
title={Non-noble plasmonic metal-based photocatalysts},
volume={122},
DOI={10.1021/acs.chemrev.1c00473},
number={11},
journal={Chemical Reviews},
author={Sayed, Mahmoud and Yu, Jiaguo and Liu, Gang and Jaroniec, Mietek},
year={2022},
month={Apr},
pages={10484–10537}
}

@article{non-noble-2,
title={Opportunities and challenges for alternative nanoplasmonic metals: Magnesium and beyond}, volume={126},
DOI={10.1021/acs.jpcc.2c01944},
number={26},
journal={The Journal of Physical Chemistry C},
author={Hopper, Elizabeth R. and Boukouvala, Christina and Asselin, Jérémie and Biggins, John S. and Ringe, Emilie},
year={2022},
month={Jun},
pages={10630–10643}
}

@article{non-noble-3,
title={Aluminum as a competitive plasmonic material for the entire electromagnetic spectrum: A review},
volume={18},
DOI={10.1016/j.rio.2024.100760},
journal={Results in Optics},
author={Shukla, S. and Arora, P.},
year={2025},
month={Feb},
pages={100760}
}

@article{al-plasmonic,
title={Aluminum nanocrystals as a plasmonic photocatalyst for hydrogen dissociation},
volume={16}, DOI={10.1021/acs.nanolett.5b05149},
number={2},
journal={Nano Letters},
author={Zhou, Linan and Zhang, Chao and McClain, Michael J. and Manjavacas, Alejandro and Krauter, Caroline M. and Tian, Shu and Berg, Felix and Everitt, Henry O. and Carter, Emily A. and Nordlander, Peter and et al.},
year={2016},
month={Jan},
pages={1478–1484}}

@article{mg-plasmonic,
title={Magnesium nanoparticle plasmonics},
volume={18}, DOI={10.1021/acs.nanolett.8b00955},
number={6},
journal={Nano Letters},
author={Biggins, John S. and Yazdi, Sadegh and Ringe, Emilie},
year={2018},
month={May},
pages={3752–3758}
}

@article{cu-plasmonic,
title={Plasmon-powered chemistry with visible-light active copper nanoparticles},
volume={15},
DOI={10.1039/d4sc04806g},
number={41},
journal={Chemical Science},
author={Tyagi, Shreya and Kashyap, Radha Krishna and Dhankhar, Ankit and Pillai, Pramod P.}, year={2024},
pages={16997–17006}
}

@article{all-bulk-surface,
  title={Ab initio calculations of generalized-stacking-fault energy surfaces and surface energies for FCC metals},
  author={Wu, Xiao-Zhi and Wang, Rui and Wang, Shao-Feng and Wei, Qun-Yi},
  journal={Applied Surface Science},
  volume={256},
  number={21},
  pages={6345--6349},
  year={2010},
  publisher={Elsevier}
}

@article{pd-ferro-1,
  title={Magnetism in coaxial palladium nanowires},
  author={Stewart, DA},
  journal={Journal of applied physics},
  volume={101},
  number={9},
  year={2007},
  publisher={AIP Publishing}
}

@article{pd-ferro-2,
  title={Comment on" Magnetism in atomic-size palladium contacts and nanowires"-Reply},
  author={Delin, A and Tosatti, E and Weht, R and others},
  journal={PHYSICAL REVIEW LETTERS},
  volume={96},
  number={7},
  pages={1--1},
  year={2006}
}

@article{persson-work-func,
  title={Work function, optical absorption, and second-harmonic generation from alkali-metal atoms adsorbed on metal surfaces},
  author={Persson, BNJ and Dubois, LH},
  journal={Physical Review B},
  volume={39},
  number={12},
  pages={8220},
  year={1989},
  publisher={APS}
}

@article{plasmonic-transformation,
  title={Photochemical transformations on plasmonic metal nanoparticles},
  author={Linic, Suljo and Aslam, Umar and Boerigter, Calvin and Morabito, Matthew},
  journal={Nature materials},
  volume={14},
  number={6},
  pages={567--576},
  year={2015},
  publisher={Nature Publishing Group UK London}
}

@article{plasmonic-transformation-1,
  author  = {Linic, Suljo and Christopher, Phillip and Ingram, David B.},
  title   = {Plasmonic-metal nanostructures for efficient conversion of solar to chemical energy},
  journal = {Nature Materials},
  volume  = {10},
  number  = {12},
  pages   = {911--921},
  year    = {2011},
  doi     = {10.1038/nmat3151},
  publisher = {Nature Publishing Group}
}

@article{plasmonic-transformation-2,
  author  = {Cort{\'e}s, Emiliano and Besteiro, Lucas V. and Alabastri, Alessandro and Baldi, Andrea and Tagliabue, Giulia and Demetriadou, Angela and Narang, Prineha},
  title   = {Challenges in Plasmonic Catalysis},
  journal = {ACS Nano},
  volume  = {14},
  number  = {12},
  pages   = {16202--16219},
  year    = {2020},
  doi     = {10.1021/acsnano.0c08773},
  publisher = {American Chemical Society}
}

@article{plasmonic-transformation-3,
  title={Catalytic conversion of solar to chemical energy on plasmonic metal nanostructures},
  author={Aslam, Umar and Rao, Vishal Govind and Chavez, Steven and Linic, Suljo},
  journal={Nature Catalysis},
  volume={1},
  number={9},
  pages={656--665},
  year={2018},
  publisher={Nature Publishing Group UK London}
}

@article{plasmonic-transformation-4,
  title={The paradox of thermal vs. non-thermal effects in plasmonic photocatalysis},
  author={Verma, Rishi and Sharma, Gunjan and Polshettiwar, Vivek},
  journal={Nature Communications},
  volume={15},
  number={1},
  pages={7974},
  year={2024},
  publisher={Nature Publishing Group UK London}
}

@article{plasmonic-transformation-5,
  title={Quantifying hot carrier and thermal contributions in plasmonic photocatalysis},
  author={Zhou, Linan and Swearer, Dayne F and Zhang, Chao and Robatjazi, Hossein and Zhao, Hangqi and Henderson, Luke and Dong, Liangliang and Christopher, Phillip and Carter, Emily A and Nordlander, Peter and others},
  journal={Science},
  volume={362},
  number={6410},
  pages={69--72},
  year={2018},
  publisher={American Association for the Advancement of Science}
}

@article{plasmonic-transformation-6,
  title={Surface-plasmon-mediated hydrogenation of carbonyls catalyzed by silver nanocubes under visible light},
  author={Landry, Michael J and Gellé, Alexandra and Meng, Beryl Y and Barrett, Christopher J and Moores, Audrey},
  journal={Acs Catalysis},
  volume={7},
  number={9},
  pages={6128--6133},
  year={2017},
  publisher={ACS Publications}
}

@article{plasmonic-transformation-7,
  author  = {Brongersma, Mark L. and Halas, Naomi J. and Nordlander, Peter},
  title   = {Plasmon-induced hot carrier science and technology},
  journal = {Nature Nanotechnology},
  year    = {2015},
  volume  = {10},
  number  = {1},
  pages   = {25--34},
  doi     = {10.1038/nnano.2014.311},
  url     = {https://doi.org/10.1038/nnano.2014.311}
}

@article{thermalization-1,
  author    = {Dubi, Yonatan and Sivan, Yonatan},
  title     = {{``Hot'' electrons in metallic nanostructures---non-thermal carriers or heating?}},
  journal   = {Light: Science \& Applications},
  volume    = {8},
  pages     = {89},
  year      = {2019},
  doi       = {10.1038/s41377-019-0199-x},
  publisher = {Nature Publishing Group}
}

@article{thermalization-2,
  author    = {Sundararaman, Ravishankar and Narang, Prineha and Jermyn, Adam S. and Goddard, William A. and Atwater, Harry A.},
  title     = {Theoretical predictions for hot-carrier generation from surface plasmon decay},
  journal   = {Nature Communications},
  volume    = {5},
  pages     = {5788},
  year      = {2014},
  doi       = {10.1038/ncomms6788},
  publisher = {Nature Publishing Group}
}

@article{thermalization-3,
  author    = {Khurgin, Jacob B.},
  title     = {How to deal with the loss in plasmonics and metamaterials},
  journal   = {Nature Nanotechnology},
  volume    = {10},
  number    = {1},
  pages     = {2--6},
  year      = {2015},
  doi       = {10.1038/nnano.2014.310},
  publisher = {Nature Publishing Group}
}

@article{thermalization-4,
  author    = {Manjavacas, Alejandro and Liu, Jun G. and Kulkarni, Vikram and Nordlander, Peter},
  title     = {Plasmon-induced hot carriers in metallic nanoparticles},
  journal   = {ACS Nano},
  volume    = {8},
  number    = {8},
  pages     = {7630--7638},
  year      = {2014},
  doi       = {10.1021/nn502445f},
  publisher = {American Chemical Society}
}

@article{thermalization-5,
  author    = {Jo{\~a}o, Sim{\~a}o M. and Jin, Hanwen and Lischner, Johannes C.},
  title     = {Atomistic theory of hot-carrier relaxation in large plasmonic nanoparticles},
  journal   = {The Journal of Physical Chemistry C},
  volume    = {127},
  number    = {48},
  pages     = {23296--23302},
  year      = {2023},
  doi       = {10.1021/acs.jpcc.3c05347},
  publisher = {American Chemical Society}
}

@article{thermalization-6,
  author  = {Saavedra, J. R. M. and Asenjo-Garcia, Ana and Garc{\'i}a de Abajo, F. Javier},
  title   = {Hot-Electron Dynamics and Thermalization in Small Metallic Nanoparticles},
  journal = {ACS Photonics},
  year    = {2016},
  volume  = {3},
  number  = {9},
  pages   = {1637--1646},
  doi     = {10.1021/acsphotonics.6b00217},
  url     = {https://doi.org/10.1021/acsphotonics.6b00217}
}

@article{kale,
  author  = {Kale, Matthew J. and
             Avanesian, Talin and
             Xin, Hongliang and
             Yan, Jun and
             Christopher, Phillip},
  title   = {Controlling Catalytic Selectivity on Metal Nanoparticles
             by Direct Photoexcitation of Adsorbate--Metal Bonds},
  journal = {Nano Letters},
  year    = {2014},
  volume  = {14},
  number  = {9},
  pages   = {5405--5412},
  doi     = {10.1021/nl502571b},
  url     = {https://doi.org/10.1021/nl502571b}
}

@article{lorentz-used-1,
  title={Chemical interface damping revealed by single-particle absorption spectroscopy},
  author={Yuan, Tinglian and Guo, Xiaofei and Lee, Stephen Anthony and Brasel, Sadie and Chakraborty, Amrita and Masiello, David J and Link, Stephan},
  journal={ACS nano},
  volume={19},
  number={10},
  pages={10277--10288},
  year={2025},
  publisher={ACS Publications}
}

@article{lorentz-used-2,
  title={The role of the plasmon in interfacial charge transfer},
  author={Ostovar, Behnaz and Lee, Stephen A and Mehmood, Arshad and Farrell, Kieran and Searles, Emily K and Bourgeois, Briley and Chiang, Wei-Yi and Misiura, Anastasiia and Gross, Niklas and Al-Zubeidi, Alexander and others},
  journal={Science Advances},
  volume={10},
  number={27},
  pages={eadp3353},
  year={2024},
  publisher={American Association for the Advancement of Science}
}

@article{dielec-1,
  title={Optical constants of the noble metals},
  author={Johnson, Peter B and Christy, R-WJPrB},
  journal={Physical review B},
  volume={6},
  number={12},
  pages={4370},
  year={1972},
  publisher={APS}
}

@article{dielec-2,
  title={Plasmonic films can easily be better: rules and recipes},
  author={McPeak, Kevin M and Jayanti, Sriharsha V and Kress, Stephan JP and Meyer, Stefan and Iotti, Stelio and Rossinelli, Aurelio and Norris, David J},
  journal={ACS photonics},
  volume={2},
  number={3},
  pages={326--333},
  year={2015},
  publisher={ACS Publications}
}

@article{dielec-3,
  title={Optical constants of transition metals: Ti, v, cr, mn, fe, co, ni, and pd},
  author={Johnson, PB and Christy, RW},
  journal={Physical review B},
  volume={9},
  number={12},
  pages={5056},
  year={1974},
  publisher={APS}
}

@article{dielec-4,
  title={Exploring the dielectric function of platinum},
  author={Tselin, Andrey and Pogodaeva, Mariia and Levchenko, Sergey and Kalmykov, Aleksey and Garbuzov, Konstantin and Smirnov, Alexander and Drachev, Vladimir},
  journal={Physical Review B},
  volume={110},
  number={19},
  pages={195130},
  year={2024},
  publisher={APS}
}

@article{dielec-5,
  title={Optical, structural, and functional properties of highly reflective and stable iridium mirror coatings for infrared applications},
  author={Schmitt, Paul and Felde, Nadja and Döhring, Thorsten and Stollenwerk, Manfred and Uschmann, Ingo and Hanemann, Kevin and Siegler, Marie and Klemm, Georg and Gratzke, Nancy and Tünnermann, Andreas and others},
  journal={Optical Materials Express},
  volume={12},
  number={2},
  pages={545--559},
  year={2022},
  publisher={Optica Publishing Group}
}

@article{dielec-6,
  title={Optical investigation of the electronic structure of bulk Rh and Ir},
  author={Weaver, JH and Olson, CG and Lynch, David W},
  journal={Physical Review B},
  volume={15},
  number={8},
  pages={4115},
  year={1977},
  publisher={APS}
}

@book{dielec-7,
  title={Handbook of optical constants of solids},
  author={Palik, Edward D},
  volume={3},
  year={1998},
  publisher={Academic press}
}

@article{dielec-8,
  title={Refractiveindex. info database of optical constants},
  author={Polyanskiy, Mikhail N},
  journal={Scientific Data},
  volume={11},
  number={1},
  pages={94},
  year={2024},
  publisher={Nature Publishing Group UK London}
}

@article{co-adsorption-1,
  title={Crystallographic study of interaction between adspecies on metal surfaces},
  author={Over, Herbert},
  journal={progress in surface science},
  volume={58},
  number={4},
  pages={249--376},
  year={1998},
  publisher={Elsevier}
}

@article{co-adsorption-2,
  title={CO adsorption energies on metals with correction for high coordination adsorption sites--A density functional study},
  author={Abild-Pedersen, Frank and Andersson, Martin P},
  journal={Surface Science},
  volume={601},
  number={7},
  pages={1747--1753},
  year={2007},
  publisher={Elsevier}
}

@article{co-adsorption-3,
  title={CO adsorption on close-packed transition and noble metal surfaces: trends from ab initio calculations},
  author={Gajdo{\v{s}}, Marek and Eichler, Andreas and Hafner, J{\"u}rgen},
  journal={Journal of Physics: Condensed Matter},
  volume={16},
  number={8},
  pages={1141--1164},
  year={2004}
}

@article{al-plasmon-1,
  title={Aluminium plasmonics},
  author={G{\'e}rard, Davy and Gray, Stephen K},
  journal={Journal of Physics D: Applied Physics},
  volume={48},
  number={18},
  pages={184001},
  year={2015},
  publisher={IOP Publishing}
}

@article{al-plasmon-2,
  title={Plasmon effects in electron energy loss and gain spectra in aluminium},
  author={Powell, BD and Woodruff, DP},
  journal={Surface Science},
  volume={33},
  number={3},
  pages={437--444},
  year={1972},
  publisher={Elsevier}
}

@article{al-plasmon-3,
  title={Energy and lifetime of surface plasmon from first-principles calculations},
  author={Silkin, VM and Chulkov, EV},
  journal={Vacuum},
  volume={81},
  number={2},
  pages={186--191},
  year={2006},
  publisher={Elsevier}
}

@article{workfunc-values,
  title={Recommended values of clean metal surface work functions},
  author={Derry, Gregory N and Kern, Megan E and Worth, Eli H},
  journal={Journal of vacuum science \& technology A},
  volume={33},
  number={6},
  year={2015},
  publisher={AIP Publishing}
}

@article{gold-tio2-persson-2-link,
  title={The role of the plasmon in interfacial charge transfer},
  author={Ostovar, Behnaz and Lee, Stephen A and Mehmood, Arshad and Farrell, Kieran and Searles, Emily K and Bourgeois, Briley and Chiang, Wei-Yi and Misiura, Anastasiia and Gross, Niklas and Al-Zubeidi, Alexander and others},
  journal={Science Advances},
  volume={10},
  number={27},
  pages={eadp3353},
  year={2024},
  publisher={American Association for the Advancement of Science}
}

@article{naldoni-paper,
  title={Platinum is a Photocatalyst: Large Visible-Light Quantum Efficiency Revealed},
  author={Henrotte, Olivier and Wong, Yin Chak and Edwards, Jordan and Jo{\~a}o, Sim{\~a}o Meneses and Kment, {\v{S}}t{\v{e}}p{\'a}n and Cort{\'e}s, Emiliano and Lischner, Johannes and Naldoni, Alberto},
  journal={arXiv preprint arXiv:2607.05172},
  year={2026}
}

\end{document}